\documentclass[12pt,showkeys,amsmath,amssymb]{revtex4}
\usepackage{amsmath,amsfonts,amsthm,amscd,amssymb,latexsym}
\usepackage{bm}
\usepackage{dcolumn}
\usepackage{graphicx}
\usepackage{epstopdf}
\usepackage{color}
\usepackage{epsf}
\usepackage{epsfig}
\usepackage{graphicx, epic, eepic, color}
\usepackage{soul}
\usepackage{tikz}
\usetikzlibrary{calc}
\usetikzlibrary{decorations.markings}
\usetikzlibrary{shadings, calc, decorations.markings}

\def\beq{\begin{equation}}
\def\eeq{\end{equation}}

\begin{document}

\title{Nonlocal Gravity: Modification of Newtonian Gravitational Force in the Solar System}

\author{Mahmood \surname{Roshan}$^{1, 2}$}
\email{mroshan@um.ac.ir}
\author{Bahram \surname{Mashhoon}$^{2,3}$}
\email{mashhoonb@missouri.edu}

\affiliation{$^1$Department of Physics, Faculty of Science, Ferdowsi University of Mashhad, P.O. Box 1436, Mashhad, Iran \\
$^2$School of Astronomy, Institute for Research in Fundamental Sciences (IPM), P. O. Box 19395-5531, Tehran, Iran\\
$^3$Department of Physics and Astronomy, University of Missouri, Columbia, Missouri 65211, USA\\
}

\begin{abstract}
Nonlocal gravity (NLG) is a classical nonlocal generalization of Einstein's theory of gravitation developed in close analogy with the nonlocal electrodynamics of media. It appears that the nonlocal aspect of the universal gravitational interaction could simulate dark matter.   Within the Newtonian regime of NLG, we investigate the deviation of the gravitational force from the Newtonian inverse square law as a consequence of the existence of the effective dark matter. In particular, we work out the magnitude of this deviation in the solar system out to 100 astronomical units. Moreover, we give an improved lower limit for the short-range parameter of the reciprocal kernel of NLG. 
\end{abstract}

\keywords{Nonlocal  gravity, Solar System, Deviations from Newton's law of gravity}

\maketitle

\section{Introduction}

The nonlocal treatment of material media in physics has a long history~\cite{Poi, Liou, Hop}. In its present form in electrodynamics, in particular, history dependence is taken into account in the constitutive properties of atomic media; that is, one retains the basic equations of Maxwell involving the electromagnetic fields $(\mathbf{E}, \mathbf{B})$ and their excitations in the medium $(\mathbf{D}, \mathbf{H})$, but the constitutive connections between these fields become nonlocal due to history dependence. The resulting constitutive relations involve a nonlocal kernel that incorporates the atomic and molecular physics of the background medium~\cite{Jackson, L+L, HeOb}. Nonlocal gravity (NLG) is a classical nonlocal generalization of Einstein's general relativity~\cite{Einstein} that has been constructed in close formal analogy with the nonlocal electrodynamics of media~\cite{Hehl:2008eu, Hehl:2009es, BMB}. There is no medium in the gravitational case; hence,  the corresponding nonlocal kernel must ultimately be determined on the basis of observational data. The detailed physical motivation for the nonlocal extension of GR and a comprehensive treatment of NLG is contained in Ref.~\cite{BMB}. 

Einstein's general relativity (GR) can be expressed in an exact form that resembles Maxwell's electrodynamics. Indeed, there is a well-known teleparallel equivalent of general relativity (TEGR), which is the gauge theory of the group of spacetime translations~\cite{Cho}. Therefore, TEGR, though nonlinear, is formally analogous to electrodynamics and can be rendered nonlocal via history-dependent constitutive relations as in the nonlocal electrodynamics of media. In the resulting theory of nonlocal gravity (NLG), there is a fundamental preferred frame field in spacetime; moreover, the gravitational field is locally defined, but satisfies partial integro-differential field equations. Nonlocal gravity theory employs an extended geometric framework involving both the Riemannian curvature of spacetime and the Weitzenb\"ock torsion of the preferred frame field. Furthermore, NLG is consistent with the universality of gravitational interaction and the principle of equivalence of inertial and gravitational masses that has strong observational support. Indeed, free test particles and null rays follow geodesics of the spacetime geometry, which ensures the universality of free fall within the framework of classical physics. The only known exact solution of the field equations of NLG is the trivial solution, namely,  Minkowski spacetime in the absence of gravity. Thus far, the nonlinearity of NLG has prevented finding exact solutions for strong-field regimes such as those involving black holes or cosmological models~\cite{Bini:2016phe}. However, linearized NLG and its Newtonian limit have been extensively studied~\cite{BMB, Mashhoon:2019jkq}. 

Heuristically, NLG theory involves a kind of spatial and temporal average of the gravitational field over the past; that is, the gravitational memory of past events survives in some form in the field equations of NLG. When one writes the field equations of NLG in the same way as GR field equations, one finds that the source term contains,  in addition to the standard symmetric energy-momentum tensor of matter, certain purely nonlocal gravity terms as well. These nonlocal gravity source terms, within the Newtonian regime of the NLG theory, help this theory recover the purely phenomenological Tohline-Kuhn modified gravity explanation of   the ``flat" rotation curves of spiral galaxies~\cite{Toh,Kuhn,Bek}. That is, the nonlocal aspect of universal gravitation appears to simulate dark matter. Therefore, it is natural to interpret the extra nonlocal gravity source terms of NLG in terms of nonlocally-induced effective dark matter, which would be a remnant of the past gravitational events. In the Newtonian regime of NLG, which is briefly described in the next section, such a memory becomes instantaneous and reduces to a spatial average with a time-independent universal kernel.   What is now considered dark matter in astrophysics and cosmology may indeed be the manifestation of the nonlocal component of the universal gravitational interaction~\cite{BMB, Roshan:2021ljs, RoMa, Roshan:2022zov}. 

In nonlocal electrodynamics, the kernel is based on the quantum physics of the medium; however, there is no medium in NLG. In this case, we must rely on astronomical data for the determination of the kernel. On the other hand, in conformity with the nonlocal electrodynamics of media, there is no Lagrangian for NLG. In fact, in Ref.~\cite{Hehl:2009es} an action principle was formulated for linearized NLG that included a time-asymmetric nonlocal kernel in connection with the past history of the gravitational field. However, the nonlocal field equation that resulted from the variation of the action involved a time-symmetric kernel that violated causality. It seems that an action principle for NLG is in conflict with causality.  

Within the Newtonian regime of NLG, the properties of effective dark matter and its distinguishing features in comparison with the standard dark matter paradigm have been the subject of recent investigations~\cite{Roshan:2021ljs, RoMa, Roshan:2022zov}. The main purpose of present work is to describe the general attributes of the gravitational force within the Newtonian regime of nonlocal gravity and study in detail its deviation from the Newtonian inverse square force law within the solar system.

\subsection{Nonlocal Newtonian Gravity}

The gravitational field equation of NLG reduces in the Newtonian limit to the nonlocal Poisson equation
\begin{equation}\label{I1}
\nabla^2\Phi (\mathbf{x}) + \int \chi(\mathbf{x}-\mathbf{y}) \nabla^2\Phi (\mathbf{y})\,d^3y = 4\pi G\,\rho (\mathbf{x})\,,
\end{equation}
where  $\chi$ is the universal kernel of NLG in the Newtonian regime.  Here, $\rho(\mathbf{x})$ is the density of matter in a Cartesian system of coordinates, $\mathbf{x} = (x, y, z)$, $G$ is Newton's constant of gravitation and $\Phi(\mathbf{x})$ is the corresponding gravitational potential.  Regarding Eq.~\eqref{I1}, we note that temporal retardation vanishes and an average over the past reduces in the limit of instantaneous connection to a spatial average when one formally lets $c \to \infty$. For the sake of simplicity, we have suppressed the possibility that $\Phi$ and $\rho$  could in principle depend upon the instantaneous temporal coordinate $t$. 
Under physically reasonable mathematical conditions, it is possible to express Eq.~\eqref{I1} in its reciprocal form 
\begin{equation}\label{I2}
4\pi G\,\rho (\mathbf{x}) +  \int q(\mathbf{x}-\mathbf{y}) [4\pi G\, \rho(\mathbf{y})]\,d^3y = \nabla^2\Phi (\mathbf{x})\,,
\end{equation}
where $q$ is the reciprocal kernel. This is the Poisson equation with an extra source term, namely,  
\begin{equation}\label{I3}
\nabla^2\Phi = 4\pi G\,(\rho+\rho_D)\,, \qquad \rho_D(\mathbf{x})=\int q(\mathbf{x}-\mathbf{y}) \rho(\mathbf{y})\,d^3y\,.
\end{equation}
It is natural to interpret $\rho_D$ as the nonlocally-induced density of effective dark matter. It is given by the convolution (``folding") of the reciprocal kernel $q$ with the density of matter $\rho$.  Therefore, the existence and distribution of effective dark matter in NLG is intimately connected with the material source and its distribution. 

The reciprocal kernel $q$ must be determined on the basis of observational data. On the other hand, NLG in the Newtonian regime requires that  $q(\mathbf{x})$ be absolutely integrable as well as square integrable over all space~\cite{BMB, Chicone:2011me}. Within the Tohline-Kuhn modified gravity scheme~\cite{Toh,Kuhn,Bek}, which NLG recovers in the Newtonian regime of the theory,  the flat rotation curves of the spiral galaxies had already led to the introduction of the spherically symmetric Kuhn kernel 
\begin{equation}\label{I4}
\frac{1}{4\pi \lambda_0}\,\frac{1}{|\mathbf{x}-\mathbf{y}|^2}\,,
\end{equation}
where $\lambda_0 \sim 1$~kpc is the basic galactic length scale in this approach.  To implement the mathematical requirements of NLG regarding $q$, we introduce two new length scales $a_0$ and $\mu_0^{-1}$ to moderate the short and long distance behaviors of the Kuhn kernel, respectively. This then leads to two possible simple positive spherically symmetric functions for $q$ that have been studied in detail~\cite{BMB}. These functions, which are integrable as well as square integrable,  are  
\begin{equation}\label{I5}
 q_1 (r) = \frac{1}{4\pi \lambda_0} \,\frac{1+\mu_0\, (a_0+r)}{r\,(a_0 + r)}\,e^{-\mu_0\,r}\,, \qquad q_2 (r) = \frac{r}{a_0 + r}\,q_1(r)\,,
\end{equation}  
where  Kuhn's kernel is recovered for $a_0 = \mu_0 = 0$ and $r = |\mathbf{x}-\mathbf{y}|$. For $a_0 = 0$,  $q_1 = q_2 = q_0$,
\begin{equation}\label{I6}
 q_0 (r) = \frac{1}{4\pi \lambda_0} \,\frac{1+\mu_0\,r}{r^2}\,e^{-\mu_0\,r}\,, 
\end{equation}
which is integrable but not square integrable. In fact, 
\begin{equation}\label{I7}
4\,\pi\,   \int_0^r  q_0(s) s^2\, ds = \alpha_0\,\left[1- (1+\tfrac{1}{2}\,\mu_0r)\,e^{-\mu_0r}\right]\,, \qquad \alpha_0 := 2/(\lambda_0\,\mu_0)\,.
\end{equation}
The integral of $q_0$ over all space is the dimensionless constant $\alpha_0$ and similarly, 
\begin{equation}\label{I8}
4\,\pi\,\int_0^\infty q_i(s) s^2\,ds = \alpha_0\, w_i\,, \quad w_1= 1- \frac{1}{2}\zeta_0\,e^{\zeta_0} E_1(\zeta_0)\,, \quad w_2 = 1- \zeta_0\,e^{\zeta_0} E_1(\zeta_0)\,,
\end{equation}
where 
\begin{equation}\label{I9}
\zeta_0 := a_0\mu_0\,
\end{equation}
is another dimensionless constant. Here,  $E_1(x)$ for $x > 0$ is the \emph{exponential integral function}~\cite{A+S}, namely,   
\begin{equation}\label{Ia} 
E_1 (x) = \int_x^\infty \frac{e^{-t}}{t}\,dt\,, \quad E^{(n)}_1(x) = \frac{d^nE_1}{dx^n}\,, \quad E_1^{(1)}(x) = - \frac{e^{-x}}{x} = -E_0(x)\,, 
\end{equation}
\begin{equation}\label{Ib} 
E_n(x) := \int_1^\infty \frac{e^{-xt}}{t^n}\,dt\,,\quad \frac{dE_n(x)}{dx} = - E_{n-1} (x)\,, \quad  x\,E_n (x) = e^{-x} - n\, E_{n+1}(x)\,, 
\end{equation}
where in the definition of $E_n(x)$,  $n = 0, 1, 2, \cdots$. 
Moreover, $q_0 > q_1 > q_2$ for any finite radial coordinate $r$. It proves useful to define 
\begin{equation}\label{I10}
\mathcal{E}_i(r)   = 4 \pi \int_0^r [q_0(s) - q_i(s)]s^2\, ds\,;
\end{equation}
then, we have for $ i = 1, 2$, 
\begin{equation}\label{I11}
4\,\pi\, \int_0^r q_i(s) s^2\,ds = -\mathcal{E}_i +  \alpha_0\,\left[1- (1+\tfrac{1}{2}\,\mu_0r)\,e^{-\mu_0r}\right]\,,
\end{equation}
where   
\begin{equation}\label{I12}
\mathcal{E}_1 (r) = \frac{1}{2} \alpha_0\, \zeta_0\,e^{\zeta_0} [E_1(\zeta_0) - E_1(\zeta_0 + \mu_0r)]\,, 
\end{equation}
\begin{equation}\label{I13}
\mathcal{E}_2 (r) = 2\, \mathcal{E}_1 (r)  -\frac{1}{2} \alpha_0\, \zeta_0\,\frac{\mu_0r}{\mu_0r+\zeta_0}e^{-\mu_0\,r}\,.
\end{equation}
The quantities $\mathcal{E}_1 (r)$ and $\mathcal{E}_2 (r)$ are positive monotonically increasing functions that start from zero at $r = 0$ and  approach $\mathcal{E}_i(\infty) = (1- w_i)\alpha_0 > 0$ as $r \to \infty$. In fact, with $w_i$, $i = 1, 2$, given by Eq.~\eqref{I8},  $\mathcal{E}_1 (\infty) = \tfrac{1}{2}\alpha_0\, \zeta_0\,e^{\zeta_0} E_1(\zeta_0)$ and $\mathcal{E}_2 (\infty) = 2\, \mathcal{E}_1 (\infty)$. 

The effective dark matter in NLG has some specific attributes that must be mentioned here.  Consider a point particle of mass $m$ that is located at $\mathbf{x}_m$; then, Eq.~\eqref{I3} implies 
\begin{equation}\label{I15}
\rho(\mathbf{x}) =  m\, \delta(\mathbf{x} - \mathbf{x}_m)\,, \qquad \rho_D(\mathbf{x}) = m \,q(\mathbf{x} - \mathbf{x}_m)\,.
\end{equation}
The reciprocal kernel $q$ is spherically symmetric by assumption; therefore, the point particle of mass $m$ is surrounded by a spherical distribution of effective dark matter of density $m\, q$ that extends to infinity and decays exponentially to zero with a decay length of $\mu_0^{-1}$. The strength of this distribution is characterized by the Tohline-Kuhn parameter $\lambda_0 \sim 1$ kpc; indeed, the effective dark matter disappears for  $\lambda_0 \to \infty$.

The net effective dark matter $m_D$ corresponding to point mass $m$ is given by
\begin{equation}\label{I16}
m_D = m\,\int q(\mathbf{x})\, d^3x\, = m\alpha_0\,w. 
\end{equation}
Extending Eq.~\eqref{I15} to a distribution of a large number of point particles, it follows that  $\rho_D$ is always finite for a Newtonian astronomical system; moreover,  $\rho_D$ traces out, albeit in an extended and diffuse manner, the form of the matter distribution. Therefore, the distribution of the effective dark matter in NLG is quite different from the standard dark matter paradigm. The fact that $\rho_D$ is the convolution of the reciprocal kernel $q$ with the density of matter $\rho$ can be used to estimate the local density of effective dark matter according to NLG in our solar neighborhood~\cite{deSalas:2020hbh}.  Moreover, it is an important property of the convolution that the total effective dark matter over all space is given by
\begin{equation}\label{I17}
M_D = \int \rho_D(\mathbf{x}) d^3x =\int q(\mathbf{x}) \,d^3x \int \rho(\mathbf{x})\,d^3x =  \alpha_0\, w\,M\,,
\end{equation}
where $M$ is the mass of the source. This is a natural generalization of Eq.~\eqref{I16} for an extended system. We can use these considerations to estimate the amount of dark matter within a given galaxy by calculating approximately the portion of $M_D$ that is confined within the boundaries defined for the galaxy~\cite{Roshan:2021ljs,Roshan:2022zov}.

\subsection{Gravitational Force in NLG}

In NLG,  as in GR, free test particles follow future-directed timelike geodesics of the spacetime metric. Naturally, in the transition from NLG to  its Newtonian regime, the speed of light  formally approaches infinity in the same way as in GR and the gravitational force on a test particle of inertial mass $m$ in a gravitational potential $\Phi(\mathbf{x})$ is thus given by the Newtonian result
\begin{equation}\label{J1}
\mathbf{F}(\mathbf{x}) = - m \nabla\Phi(\mathbf{x})\,.
\end{equation}

Imagine a point particle of mass $m$ at $\mathbf{x}_m$ as in Eq.~\eqref{I15}; in this case, Eq.~\eqref{I3} implies that the gravitational potential is due to the point mass $m$ as well as its cocoon of  effective dark matter. The latter forms a spherical distribution centered on $m$ with density $m q$. Imagine now another point particle of mass $m'$ located at $\mathbf{x}'$. The gravitational force on $m'$ due to $m$ can be calculated using Eq.~\eqref{J1}; indeed, the result is
\begin{equation}\label{J2}
\mathbf{F}_{\rm NLG}(\mathbf{x}') =  Gm' \frac{\mathbf{x}_m - \mathbf{x}'}{|\mathbf{x}_m - \mathbf{x}'|^3} \left[m+ m\int_0^{|\mathbf{x}_m - \mathbf{x}'|} 4\,\pi\,s^2 q(s) ds\right]\,,
\end{equation}
where the first term in the square brackets is simply the Newtonian result due to point mass $m$, while the second term is due to its cocoon, namely,  the spherical distribution of effective dark matter of density $mq$ surrounding $m$. We have employed Newton's shell theorem here, since $m'$ is affected only by the attractive force of the portion of the cocoon that is within a sphere centered on $m$ of radius $|\mathbf{x}_m -\mathbf{x}'|$, the distance from $m$ to $m'$. In fact, $m'$ is unaffected by the remaining mass of the cocoon as a direct consequence of Newton's shell theorem. Furthermore, this theorem states that the spherical portion of the effective dark matter of radius $|\mathbf{x}_m -\mathbf{x}'|$ acts as though it were concentrated at $\mathbf{x}_m$.     It follows that
\begin{equation}\label{J3}
\mathbf{F}_{\rm NLG}(\mathbf{x}') =  Gm'm \frac{\mathbf{x}_m - \mathbf{x}'}{|\mathbf{x}_m - \mathbf{x}'|^3} \left[1 + \Delta(|\mathbf{x}_m -\mathbf{x}'|) \right]\,,
\end{equation}
where Eq.~\eqref{I11} implies
\begin{equation}\label{J4}
\Delta (r)  = \int_0^r 4\,\pi\,s^2 q(s)\, ds = - \mathcal{E}(r)+\alpha_0\,[1- (1+\tfrac{1}{2}\,\mu_0r)\,e^{-\mu_0r}]\,.
\end{equation}
Here,  $\Delta$ contains the  contribution of the effective dark matter and $\mathcal{E}(r)$, which vanishes when $a_0 = 0$,  has been defined in Eqs.~\eqref{I12} and~\eqref{I13}. We note that $\Delta(r)$ is a positive monotonically increasing function that starts from zero at $r=0$ and asymptotically approaches $\Delta(\infty)= \alpha_0 w$ in accordance with Eq.~\eqref{I8}. 

We can now extend $\mathbf{F}_{\rm NLG}$ to a matter distribution. That is, the NLG force on a point mass $m$ at $\mathbf{x}$ due to a matter  distribution with density $\rho(\mathbf{x})$ is given by 
\begin{equation}\label{J5}
 \mathcal{F}_{\rm NLG}(\mathbf{x})=-G m\int \left[1 + \Delta(|\mathbf{x}-\mathbf{y}|) \right]\,\frac{\mathbf{x}-\mathbf{y}}{|\mathbf{x}-\mathbf{y}|^3}\rho(\mathbf{y})\,d^3y\,.
\end{equation}
Let us first imagine that we are in the exterior of a finite astronomical source. As we move away from the source, the gravitational attraction increases due to the effective dark matter and eventually the Newtonian  inverse square law is recovered; that is, very far from the source, $|\mathbf{x} - \mathbf{y}| \gg \mu_0^{-1}$ and we find $\Delta(|\mathbf{x}-\mathbf{y}|)\approx \alpha_0 w$. Hence, Eq.~\eqref{J5} reduces to the Newtonian result
\begin{equation}\label{J6}
 \mathcal{F}_{\rm NLG}(\mathbf{x})\approx -G(1+\alpha_0 w) m\int \frac{\mathbf{x}-\mathbf{y}}{|\mathbf{x}-\mathbf{y}|^3}\rho(\mathbf{y})\,d^3y\,,
\end{equation}
except that Newton's constant $G$ has now been replaced in effect by $G(1+\alpha_0 w)$. For $|\mathbf{x}| \gg |\mathbf{y}|$, we have to lowest order  in $|\mathbf{y}|/|\mathbf{x}|$,
\begin{equation}\label{J7}
 \mathcal{F}_{\rm NLG}(\mathbf{x})\approx -G(1+\alpha_0 w) mM \frac{\mathbf{x}}{|\mathbf{x}|^3}\,,
\end{equation}
where $M$ is the mass of the source and $(1+\alpha_0 w)M$ is its mass plus its associated net effective dark matter. For a galactic source, the corresponding rotation curve, where the attractive acceleration of gravity equals the centripetal acceleration of circular motion,  is then essentially Newtonian very far from the galaxy. This consequence of NLG appears to be consistent with the conclusion of a recent study of the far-flung rotation data about Milky Way and M31~\cite{Dai:2022had}. 

For the sake of completeness, let us mention that the gravitational potential in NLG for a matter distribution of density $\rho$ is given by
\begin{equation}\label{J8}
^{\rm NLG} \Phi_i(\mathbf{x})=-G\,\int \Big[1 -\mathcal{E}_i(\infty)+\alpha_0 (1- e^{-\mu_0 r})+\frac{r}{\lambda_0}(1+i\frac{a_0}{r}) e^{\zeta_0} E_1(\zeta_0 + \mu_0 r)\Big]\,\frac{\rho(\mathbf{y})}{|\mathbf{x}-\mathbf{y}|}\,d^3y\,,
\end{equation}
where $r=|\mathbf{x}-\mathbf{y}|$ and $i=1, 2$, corresponding to $q_1$ and $q_2$, respectively. In fact, one can show explicitly using Eq.~\eqref{J5} that $\mathcal{F}_{\rm NLG}(\mathbf{x}) =  -  m\,\nabla ^{\rm NLG}\Phi(\mathbf{x})$.

In Eq.~\eqref{I3} for the effective dark matter density $\rho_D$, $q=0$ leads to $\rho_D = 0$ and we recover Poisson's equation of Newtonian gravity that is a consequence of the inverse square law. Therefore, we expect that NLG leads to deviations from the inverse square law as a result of the existence of the effective dark matter. On small laboratory scales, there has been great progress in verifying Newton's law of gravitation down to a distance of about 50 $\mu$m and efforts are under way to explore even smaller distances~\cite{Adelberger:2003zx, Tan:2016vwu, Tan:2020vpf, Lee:2020zjt, Baeza-Ballesteros:2021tha, Du:2022veu}. For the sake of completeness, we mention here that the leading quantum correction to the Newtonian gravitational potential, $-GM/r$, due to mass $M$ is given by~\cite{dePaulaNetto:2021axj} 
\begin{equation}\label{J9}
-\frac{GM}{r}\left(1+ \frac{17}{ 20 \pi} \frac{L_P^2}{r^2}\right)\,,
\end{equation}
where $L_P = (\hbar G/c^3)^{1/2} \approx 10^{-33}$ cm is the Planck length. In NLG, $G$ is a constant of nature as well, since there is no firm observational evidence in support of a variable $G$.
Meanwhile, terrestrial experiments regarding Newton's law and the measurement of Newton's constant of gravitation $G$ continue at present~\cite{Dai:2021jnl, Bhagvati:2021ksp}. On the other hand,  interesting experiments have been proposed recently to measure directly deviations from the inverse square law of gravity in the solar system out to 100 astronomical units (AU) and beyond~\cite{Buscaino:2015fya, Feldman:2016pws, Berge:2019zjj, Zwick:2022hbe}. In any case, modifications are expected due to the presence of dark matter~\cite{Belbruno:2022hsm}. Finally, on larger galactic scales, we expect to determine the reciprocal kernel $q$ from the rotation curves of spiral galaxies. By fitting the predictions of NLG with observational data, we hope to find $a_0$, $\alpha_0$ and $\mu_0$. However, in connection with the short-distance parameter $a_0$, reliable rotational data is not available very close to the center of the galactic bulge. It therefore appears that we could ignore the short-range parameter $a_0$ when we compare the predictions of NLG with the rotation curves of spiral galaxies.  In this way, $\alpha_0$ and $\mu_0$ have been determined; in fact,  
\begin{equation}\label{J10}
\alpha_0 = 10.94 \pm 2.56\,,\quad \mu_0 = 0.059 \pm 0.028~{\rm kpc}^{-1}\,, \quad  \lambda_0 = \frac{2}{\alpha_0\,\mu_0} \approx 3\pm 2~{\rm kpc}\,
\end{equation}
based on the observational data regarding nearby galaxies and clusters of galaxies~\cite{Rahvar:2014yta}. We should mention the possibility that the three parameters of the reciprocal kernel could in principle depend upon cosmological time; here, however, we assume that they refer to the present cosmological epoch~\cite{BMB}.
Regarding the observational determination of $a_0$, we note that a lower bound for $a_0$, namely, $a_0 \gtrsim 10^{15}$~cm, or about 100 AU, was calculated in 2015 using the solar system data available at that time in connection with the perihelion precession rate of Saturn~\cite{Chicone:2015coa}. The implications of NLG for the solar system crucially depend on the value of $a_0$ and we will employ current solar system data to improve the lower bound on $a_0$ in Section IV.
 
The aim of this paper is to present a general discussion of the gravitational force within the framework of NLG; in particular, we henceforth focus on deviations from the inverse square law of gravitation in the solar system out to 100 AU and beyond, which could be measurable in principle via future experiments~\cite{Buscaino:2015fya, Feldman:2016pws, Berge:2019zjj, Zwick:2022hbe}.  

\section{NLG Force in the Solar System}  

Let us start with Eq.~\eqref{J5} for the force of gravity according to NLG due to a compact object such as a star. We are interested in the force in the exterior of the source, $|\mathbf{x}| > |\mathbf{y}|$, but with
\begin{equation}\label{J11}
|\mathbf{x} - \mathbf{y}| \ll \mu_0^{-1} \approx 17\, {\rm kpc}\,.
\end{equation} 
This is the situation of interest in connection with solar system experimental proposals contained in Refs.~\cite{Buscaino:2015fya, Feldman:2016pws, Berge:2019zjj, Zwick:2022hbe}. In view of Eq.~\eqref{J11},  we want to expand $\Delta(r)$ in Eq.~\eqref{J5} in powers of $\mu_0 r \ll 1$.  To this end, we start with Eq.~\eqref{J4}, where $\mathcal{E}(r)$ is given by Eqs.~\eqref{I12} and~\eqref{I13}.  For $\mathcal{E}_1$ in  Eq.~\eqref{I12}, let us expand $E_1(\zeta_0 + \mu_0r)$  in a Taylor series about $E_1(\zeta_0)$ to get 
\begin{equation}\label{J12} 
\mathcal{E}_1(r) = -\frac{1}{2} \alpha_0 \zeta_0 e^{\zeta_0} \sum_{n = 1}^{\infty} \frac{(\mu_0r)^n}{n!} E^{(n)}_1(\zeta_{0})\,.  
\end{equation} 
From the definition of the exponential integral function in Eq.~\eqref{Ia}, we find for $n = 0, 1, 2, \cdots$, 
\begin{equation}\label{J13} 
E^{(n+1)}_1 (x) = (-1)^{n+1} n! \,\frac{e^{-x}}{x^{n+1}}\,W_{n}(x)\,,
\end{equation}
where $W_n$ is a polynomial of degree $n$ defined by
\begin{equation}\label{Ja} 
 W_n (x) := \sum_{k=0}^n \frac{x^k}{k!}\,.
\end{equation}
For $n \to \infty$, $W_n (x) \to e^x$. Furthermore, in terms of these polynomials, we have for $\mathcal{E}_2(r)$ in Eq.~\eqref{I13},
\begin{equation}\label{J14} 
 \frac{a_0}{a_0 + r}e^{-\mu_0 r} = \sum_{n=0}^{\infty}(-1)^n W_n(\zeta_0) \left(\frac{r}{a_0}\right)^n\,,
\end{equation}
where $r<a_0$ by assumption.  Then, putting these results together, we have
\begin{equation}\label{J15} 
 \mathcal{E}_1(r) = \frac{r}{\lambda_0} \sum_{n=0}^{\infty}\frac{(-1)^n}{n+1} W_n(\zeta_0) \left(\frac{r}{a_0}\right)^n\,
\end{equation}
and
\begin{equation}\label{J16} 
 \mathcal{E}_2(r) = \frac{r}{\lambda_0} \sum_{n=0}^{\infty}(-1)^{n+1}\,\frac{n-1}{n+1} \,W_n(\zeta_0) \left(\frac{r}{a_0}\right)^n\,.
\end{equation}
Remarkably, our expansions in powers of $\mu_0 r \ll 1$, where $\mu_0^{-1} \approx 17$ kpc is the long-range parameter of the reciprocal kernel $q$, end up being in effect expansions in powers of $r/a_0$, where $a_0$ is the short-range parameter of $q$.  Writing Eq.~\eqref{J4} in the form
\begin{equation}\label{J17} 
 \Delta(r) = -\mathcal{E}(r) + \frac{r}{\lambda_0} \sum_{n=0}^{\infty}(-1)^{n+1}\,\frac{n-1}{(n+1)!}\, \zeta_0^n \left(\frac{r}{a_0}\right)^n\,,
\end{equation}
we get expansions
\begin{equation}\label{J18} 
 ^{(1)}\Delta(r) = \frac{r}{\lambda_0} \sum_{n=1}^{\infty}\frac{(-1)^{n+1}}{n+1} \left[W_n(\zeta_0)+\frac{n-1}{n!}\zeta_0^n\right] \left(\frac{r}{a_0}\right)^n\,
\end{equation}
and
\begin{equation}\label{J19} 
^{(2)}\Delta(r) = \frac{r}{\lambda_0} \sum_{n=1}^{\infty}(-1)^{n+1}\frac{n}{n+2} \,W_n(\zeta_0)\, \left(\frac{r}{a_0}\right)^{n+1}\,,
\end{equation}
for  $q_1$ and $q_2$, respectively.

Let us write $\Delta$ in Eq.~\eqref{J5}  as
\begin{equation}\label{J20}
 \Delta(|\mathbf{x}-\mathbf{y}|) = \sum_{n = 2}^{\infty} \Delta_n(\zeta_0)\,\frac{|\mathbf{x}-\mathbf{y}|^n}{\lambda_0 a_0^{n-1}}\,,
\end{equation} 
where $\Delta_n(\zeta_0)$ are dimensionless functions of $\zeta_0 = \mu_0 a_0$. 
Comparing  relations~\eqref{J18}--\eqref{J19} with Eq.~\eqref{J20}, we conclude that for $n = 2, 3, 4, \cdots$, 
\begin{equation}\label{J21} 
 ^{(1)}\Delta_n(\zeta_0) = \frac{(-1)^{n}}{n} \left[W_{n-1}(\zeta_0)+\frac{n-2}{(n-1)!}\zeta_0^{n-1}\right]\,
\end{equation}
and
\begin{equation}\label{J22} 
^{(2)}\Delta_n(\zeta_0) = (-1)^{n-1}\,\frac{n-2}{n} \,W_{n-2}(\zeta_0)\,,
\end{equation}
where $W_n(\zeta_0) = \sum_{k = 0}^n \zeta_0^k/k!$. Therefore, depending on whether we use $q_1$ or $q_2$,  we find
\begin{equation}\label{J23}
^{(1)}\Delta(r)  = \frac{1}{2} (1+\zeta_0) \frac{r^2}{\lambda_0 a_0} -  \frac{1}{3}(1+\zeta_0 + \zeta_0^2) \frac{r^3}{\lambda_0 a_0^2} +  \frac{1}{4}(1+\zeta_0 + \tfrac{1}{2}\zeta_0^2 + \tfrac{1}{2}\zeta_0^3) \frac{r^4}{\lambda_0 a_0^3} - \cdots\,
\end{equation}
or
\begin{equation}\label{J24}
^{(2)}\Delta(r)  =   \frac{1}{3}(1+\zeta_0) \frac{r^3}{\lambda_0 a_0^2} -  \frac{1}{2}(1+\zeta_0 + \tfrac{1}{2}\zeta_0^2) \frac{r^4}{\lambda_0 a_0^3} + \cdots\,,
\end{equation}
respectively. We expect that these series converge for $r < a_0$. It is straightforward to calculate higher-order terms in these series, which we must substitute in Eq.~\eqref{J5} in order to calculate general NLG deviations from Newton's law of universal gravitation. This is explicitly illustrated in the rest of this section for a spherically symmetric source.

\subsection{Spherically Symmetric Distribution of Matter} 

Imagine a spherically symmetric distribution of matter confined within a sphere of finite radius $r_0$.  Each point particle of mass $m$ has an accompanying sphere of effective dark matter of density $mq$ surrounding it and the spherical symmetry of mass distribution naturally results in an effective dark matter distribution that is spherically symmetric and concentric with the matter distribution. To see this in detail, consider a Cartesian coordinate system $(x, y, z)$ with its origin at the center of the spherical source of radius $r_0$. In the convolution formula for $\rho_D(\mathbf{x})$ in Eq.~\eqref{I3}, let the orientation of the coordinate system be such that $\mathbf{x}= (0, 0, r)$ and $\mathbf{y}= (y\sin\theta\cos\phi, y\sin\theta\sin\phi, y\cos\theta)$, with no loss in generality. Then, after integration over the azimuthal angle $\phi$, we can write 
\begin{equation}\label{Ka} 
\rho_D(r)= 2\pi \int  \rho(y)\,y^2\,dy\,q(U) \sin\theta\,d\theta\,, 
\end{equation}
where 
\begin{equation}\label{K1}
U  := |\mathbf{x}-\mathbf{y}| = (r^2 -2ry \cos\theta + y^2)^{1/2}\,,\qquad  \frac{\partial U}{\partial (\cos\theta)} = -\frac{ry}{U}\,.
\end{equation}
The integration over the remaining angular coordinate is straightforward and results in  
\begin{equation}\label{K2}
\rho_D(r) = \frac{2\pi}{r}\int_0^{r_0} \rho(y)\,y\,[\Psi(r+y) - \Psi(|r-y|)]\,dy\,,
\end{equation}
where
\begin{equation}\label{K3}
\Psi(s)  = \int^s q(u) u \,du\,.
\end{equation}

Let us next calculate the gravitational force experienced by a test point particle of mass $m$ due to the concentric spherical distributions of $\rho$ and $\rho_D$. The result is simply obtained via Newton's shell theorem, namely, the attractive force of gravity is radial with magnitude $\mathcal{F}_N + \mathcal{F}_D$, where $\mathcal{F}_N$ is the strictly Newtonian part due to $\rho$; that is, 
\begin{equation}\label{K4}
\mathcal{F}_N(R) =  \frac{4\pi Gm}{R^2} \int_{0}^{R}\rho(r) r^2 \,dr\,,\quad \mathcal{F}_D(R) =  \frac{4\pi Gm}{R^2} \int_{0}^{R}\rho_D(r) r^2 \,dr\,,
\end{equation}
where $R$ is the distance from $m$ to the common center of $\rho$ and $\rho_D$. 

In Appendix A, $\rho_D(r)$ is explicitly worked out for a spherical source of constant density $\rho_0$. In this case, an exact analytic formula can be obtained for the force $\mathcal{F}_D$  due to effective dark matter. On the other hand, we are interested in the deviations of the gravitational force from the inverse square law in the solar system in connection with recent experimental proposals~\cite{Buscaino:2015fya, Feldman:2016pws, Berge:2019zjj, Zwick:2022hbe}; therefore, for practical purposes, we use the approximation scheme involved in expansions~\eqref{J23}--\eqref{J24} to derive an alternative formula for $\mathcal{F}_D(R)$ as described below.

\subsection{Alternative Formula for $\mathcal{F}_D(R)$}

Alternatively, we can use Eq.~\eqref{J5} to compute the gravitational force in NLG when the source of mass $M$ is spherically symmetric and confined within a sphere of radius $r_0$. To compute the relevant integral in  Eq.~\eqref{J5}, we employ the same approach as in the previous subsection  in order to find the gravitational force on $m$ when it is located \emph{outside} the source, namely,  $R > r_0$. The force is then along the negative $z$ direction and has magnitude $\mathcal{F}_N + \mathcal{F}_D$, where the Newtonian part is simply given by $\mathcal{F}_N = G m M/R^2$ and the effective dark matter part is given by
\begin{equation}\label{K5}
\mathcal{F}_D(R) =  2\pi Gm \int_{0}^{r_0}\rho(y) y^2 \,dy \int_{-1}^{1} \frac{\Delta (\mathcal{U})}{\mathcal{U}^3} (R - y\cos \theta)\, d(\cos\theta)\,,
\end{equation}
where  $\mathcal{U}$ is defined by
\begin{equation}\label{K6}
\mathcal{U}  := (R^2 -2Ry \cos\theta + y^2)^{1/2}\,.
\end{equation} 

In principle,  expansions~\eqref{J23} and~\eqref{J24} could be continued for  $U < a_0$ and the resulting convergent infinite series could be used in Eq.~\eqref{K5} to determine the contribution of effective dark matter to the gravitational force. That is, let us expand $\Delta(\mathcal{U})$ in powers of $\mathcal{U}$ as in Eq.~\eqref{J20}. Then,  with 
\begin{equation}\label{Kb}
R - y\cos \theta = \frac{1}{2R}(\mathcal{U}^2 +R^2-y^2)\,
\end{equation} 
and employing the same integration approach as above, we can write Eq.~\eqref{K5} in the form
\begin{equation}\label{K7}
\frac{1}{m} \mathcal{F}_D(R) =  \frac{4\pi G}{R^2} \sum_{n = 2}^{\infty} \frac{ \Delta_n(\zeta_0)}{(n^2-1)\lambda_0 a_0^{n-1}}\int_{0}^{r_0} \mathcal{I}_n(y;R)\rho(y) y^2 \,dy\,.
\end{equation}
Here, we have introduced
\begin{equation}\label{K8}
\mathcal{I}_n(y;R)  = \frac{1}{2y}\left[(nR-y)(R+y)^n - (nR+y)(R-y)^n\right]\,,
\end{equation} 
which is an even function of $y$, while $\mathcal{I}_n(y; -R)  = (-1)^n \mathcal{I}_n(y;R)$. We note that $\mathcal{I}_n(0;R) = (n^2-1)R^n$; moreover, $\mathcal{I}_n(y;0) = - y^n$ for $n=$ even, while $\mathcal{I}_n(y;0) = 0$ for $n=$ odd. Specifically, 
\begin{align}\label{K9}
\nonumber   {}&\mathcal{I}_2(y;R)  = 3R^2-y^2\,, \quad \mathcal{I}_3(y;R)  = 8R^3\,, \quad \mathcal{I}_4(y;R)  = 15R^4+ 10 R^2y^2  -y^4\,,\\
{}&\mathcal{I}_5(y;R)  = 24R^5+ 40 R^3y^2\,, \quad \mathcal{I}_6(y;R)  = 35R^6+ 105 R^4y^2 + 21 R^2y^4 -y^6\,,\cdots.
\end{align} 
Plugging these expressions into Eq.~\eqref{K7}, we find that the deviation from  the Newtonian gravitational force involves the even moments of the mass density $\rho$. 

To illustrate this general approach, we next calculate the NLG modification of the gravitational force explicitly for a spherical source of \emph{constant} density $\rho_0$.

\subsection{$\mathcal{F}_D(R)$ for a Spherical Source of Constant Density}

Let us now assume that the density of matter is constant and equal to $\rho_0$ such that the mass of the source is given by 
\begin{equation}\label{L1}
M  = \frac{4 \pi}{3}\rho_0\,r_0^3\,.
\end{equation} 
Then, the integration over the source in Eq.~\eqref{K7} can be simply carried through and the result takes the form
\begin{equation}\label{L2}
\frac{1}{m} \mathcal{F}_D(R) =  \frac{GM}{\lambda_0 a_0} \left[\Delta_2 \left(1-\frac{1}{5} \frac{r_0^2}{R^2} \right) + \Delta_3 \frac{R}{a_0}+ \Delta_4 \left(1+\frac{2}{5} \frac{r_0^2}{R^2} - \frac{1}{35} \frac{r_0^4}{R^4}\right)\frac{R^2}{a_0^2}+ \cdots \right]\,,
\end{equation}
where we can now substitute $\Delta_n(\zeta_0)$ from either Eq.~\eqref{J21} or Eq.~\eqref{J22} to determine the force due to the effective dark matter depending on whether we use reciprocal kernel $q_1$ or $q_2$. Therefore, outside the source, we have the important result
\begin{equation}\label{L3}
\frac{\mathcal{F}_D(R)}{\mathcal{F}_N(R)}  =  \frac{R^2}{\lambda_0 a_0} \left[\Delta_2 \left(1-\frac{1}{5} \frac{r_0^2}{R^2} \right) + \Delta_3 \frac{R}{a_0}+ \Delta_4 \left(1+\frac{2}{5} \frac{r_0^2}{R^2} - \frac{1}{35} \frac{r_0^4}{R^4}\right)\frac{R^2}{a_0^2}+ \cdots \right]\,.
\end{equation}

As mentioned before, we are interested in the experiments~\cite{Buscaino:2015fya, Berge:2019zjj} that propose to measure deviations from Newton's inverse square law in the solar system at the level of about $10^{-7}$ out to a distance of 100 AU. Let us note that with $a_0  \gtrsim 100$ AU, the amplitude of NLG deviation from the inverse square law given by Eq.~\eqref{L3} at $R = 100$ AU is
\begin{equation}\label{L4}
\frac{R^2}{\lambda_0 a_0} \lesssim \frac{R}{\lambda_0} \approx 1.5 \times 10^{-7}\,,
\end{equation}
where $\lambda_0 \approx 3$ kpc in agreement with Eq.~\eqref{J10}. Furthermore, for the Sun, $r_0 = R_\odot \approx 5 \times 10^{-3}$ AU; therefore, beyond the orbit of the Earth,  terms of order $r_0^2/R^2$ and higher may be neglected in Eqs.~\eqref{L2}--\eqref{L3}. This means that the NLG deviation from the Newtonian acceleration of gravity beyond the orbit of the Earth can be approximated by 
\begin{equation}\label{L5}
\frac{\mathcal{F}_D(R)}{\mathcal{F}_N(R)} \approx  \frac{R^2}{\lambda_0 a_0} \sum_{n = 0}^{\infty} \Delta_{n +2}(\zeta_0) \left(\frac{R}{a_0}\right)^n\,,
\end{equation}
which is due to the effective dark matter of the Sun. General expressions for $^{(1)}\Delta_{n+2}(\zeta_0)$ and $^{(2)}\Delta_{n+2}(\zeta_0)$, $n = 0, 1, 2, \cdots$,  are given in Eqs.~\eqref{J21}--\eqref{J22}. 

The effective dark matter of the Sun affects the planetary orbits as well. The extra acceleration of gravity experienced by a planet outside the orbit of the Earth  is given by 
\begin{equation}\label{L6}
\frac{1}{m} \mathcal{F}_D(R)|_{q = q_1} \approx  \frac{GM_\odot}{\lambda_0 a_0} \left[\frac{1}{2} (1+\zeta_0) -\frac{1}{3} (1+\zeta_0 + \zeta_0^2)\frac{R}{a_0}+\frac{1}{4} (1+\zeta_0 +\tfrac{1}{2} \zeta_0^2 +\tfrac{1}{2} \zeta_0^3)\frac{R^2}{a_0^2}+ \cdots \right]\,
\end{equation}
or
\begin{equation}\label{L7}
\frac{1}{m} \mathcal{F}_D(R)|_{q = q_2} \approx  \frac{GM_\odot}{\lambda_0 a_0} \left[\frac{1}{3} (1+\zeta_0)\frac{R}{a_0} - \frac{1}{2} (1+\zeta_0 +\tfrac{1}{2} \zeta_0^2)\frac{R^2}{a_0^2}+ \cdots \right]\,,
\end{equation}
depending upon whether we use $q_1$ or $q_2$ for the reciprocal kernel of NLG. We now turn to the general implications of Eq.~\eqref{L2} for the orbits of  planets in the solar system. 

\section{NLG: Planetary Orbits in the Solar System} 

Imagine a planet of mass $m$ in orbit about a star of mass $M$ within the framework of Newtonian mechanics. This is the standard classical two-body problem in Newtonian celestial mechanics and it is possible to formulate it exactly within the Newtonian regime of NLG by taking due account of the effective dark matter of the star as well as the planet. To simplify matters, however, we write the equation of relative motion in our case as 
\begin{equation}\label{M1}
\frac{d^2 \mathbf{R}}{dt^2} + \frac{GM \mathbf{R}}{R^3} = -\mathbb{F}(R)\, \frac{ \mathbf{R}}{R}\,,
\end{equation}
where we have assumed $m \ll M$,  $M + m \approx M$ and that the force due to the planet's effective dark matter can be neglected. Here, $\mathbb{F}(R)$ is the magnitude of the perturbing function due to the effective dark matter of the star. The NLG perturbing force is central, attractive and conservative; hence, the relative orbit is planar. In the background Cartesian coordinate system $(X, Y, Z)$, let $(X, Y)$ be the orbital plane; moreover, we introduce polar coordinates $(R, \varphi)$ such that  $X = R \cos \varphi$ and $Y = R \sin\varphi$. The unperturbed relative orbit is an ellipse in the $(X, Y)$ plane with one of its foci at the origin of coordinates. That is, when we turn off the perturbation in Eq.~\eqref{M1},  $\mathbf{R} = (X, Y, Z)$ with  $Z=0$ and
\begin{equation}\label{M2}
R = \frac{\mathbb{A}_0(1-\mathbb{E}_0^2)}{1+ \mathbb{E}_0 \cos(\varphi-\mathbb{G}_0)}\,,
\end{equation}
where $\mathbb{A}_0$, $\mathbb{E}_0$ and $\mathbb{G}_0$ are the semimajor axis, eccentricity and argument of the pericenter of the unperturbed relative orbit, respectively. The unperturbed pericenter has coordinates
\begin{equation}\label{Ma}
\mathbb{A}_0(1-\mathbb{E}_0)(\cos\mathbb{G}_0, \sin\mathbb{G}_0, 0)\,.
\end{equation}
Moreover, motion along the unperturbed orbit takes place such that 
\begin{equation}\label{M3}
\frac{d\varphi}{dt}  = \frac{L_0}{R^2}\,, \qquad L_0 = [G M\mathbb{A}_0(1-\mathbb{E}_0^2)]^{1/2}\,,
\end{equation}
where $m L_0$ is the orbital angular momentum of the unperturbed relative orbit. Let $T$ be the period of the unperturbed orbit, then the Keplerian frequency $\Omega$ is given by
\begin{equation}\label{M4}
\Omega  = \frac{2 \pi}{T}\,, \qquad \Omega^2 = \frac{G M}{\mathbb{A}_0^3}\,.
\end{equation}

For the perturbed system~\eqref{M1}, the instantaneous position and velocity of the relative motion define an \emph{osculating} ellipse that is momentarily tangent to the perturbed orbit; therefore, the perturbed relative motion can be described via the evolution of the osculating ellipse given by the Lagrange planetary equations~\cite{Danby}. In the present planar case, the Lagrange planetary equations reduce to 
\begin{equation}\label{M5}
\frac{d \mathbb{A}}{dt} = -\frac{2 \mathbb{E} \sin(\varphi-\mathbb{G})}{\Omega (1-\mathbb{E}^2)^{1/2}} \,\mathbb{F}(R)\,,
\end{equation}
\begin{equation}\label{M6}
\frac{d \mathbb{E}}{dt} = \frac{1-\mathbb{E}^2}{2 \mathbb{A}\,\mathbb{E}}\, \frac{d \mathbb{A}}{dt}\,,
\end{equation}
\begin{equation}\label{M7}
\frac{d \mathbb{G}}{dt} =  \frac{(1-\mathbb{E}^2)^{1/2}}{ \mathbb{A}\, \mathbb{E}\,\Omega}\,\mathbb{F}(R) \cos(\varphi-\mathbb{G})\,.
\end{equation}

To reveal the long-term behavior of the perturbed Keplerian system~\eqref{M1}, we average over the ``fast" Keplerian motion of frequency $\Omega$. In our approximation scheme, we define the average of a function $f$ by
\begin{equation}\label{M8}
< f >  = \frac{1}{T}\,\int_0^T f dt = \frac{1}{2\pi}(1-\mathbb{E}_0^2)^{3/2}\int_0^{2\pi} \frac{f(\varphi) d\varphi}{[1+ \mathbb{E}_0 \cos(\varphi-\mathbb{G}_0)]^2}\,, 
\end{equation}
where Eqs.~\eqref{M2}--\eqref{M4} for the unperturbed orbit have been employed. Let us now write
\begin{equation}\label{M9}
\mathbb{F}(R)  = \sum_{n=0}^{\infty} \mathbb{F}_n R^n\,, \qquad \mathbb{F}_n = \frac{GM}{\lambda_0 a_0^{n+1}} \Delta_{n +2}\,, 
\end{equation}
using Eq.~\eqref{L5}. Here, $R$ is given by Eq.~\eqref{M2} for the unperturbed orbit. Averaging Eqs.~\eqref{M5}--\eqref{M6}, we find 
\begin{equation}\label{M10}
\left \langle \frac{d \mathbb{A}}{dt} \right \rangle  = \left \langle \frac{d \mathbb{E}}{dt} \right \rangle = 0\,,
\end{equation}
which means that on average the shape and size of the orbit remain unchanged. Moreover, the Keplerian ellipse slowly precesses and the rate of precession is given by
\begin{equation}\label{M11}
\left \langle \frac{d \mathbb{G}}{dt} \right \rangle  = -\frac{1}{\mathbb{E}_0\,\Omega} \sum_{n=0}^{\infty}\mathbb{A}_0^{n-1} (1-\mathbb{E}_0^2)^{n+2}\mathbb{F}_n\,\mathbb{I}_n\,,
\end{equation}
where 
\begin{equation}\label{M12}
\mathbb{I}_n = -\frac{1}{2\pi} \int_{\eta_0}^{\eta_0+2\pi} \frac{\cos \eta\,d\eta}{(1+ \mathbb{E}_0 \cos \eta)^{n+2}}\,.
\end{equation}
Here, $\eta_0$ is a constant. By taking derivatives  of the integrals below with respect to the eccentricity $\mathbb{E}_0$, 
\begin{equation}\label{M13}
\frac{1}{2\pi} \int_{\eta_0}^{\eta_0+2\pi} \frac{d\eta}{1+ \mathbb{E}_0 \cos \eta} = \frac{1}{(1- \mathbb{E}_0^2)^{1/2}}\,, \quad \frac{1}{2\pi} \int_{\eta_0}^{\eta_0+2\pi} \frac{d\eta}{(1+ \mathbb{E}_0 \cos \eta)^{2}} = \frac{1}{(1- \mathbb{E}_0^2)^{3/2}},
\end{equation}
we can evaluate $\mathbb{I}_0$ and $\mathbb{I}_1$. More generally, 
\begin{equation}\label{M14}
\mathbb{E}_0\,\frac{\partial \mathbb{I}_n}{\partial \mathbb{E}_0}  = (n+2) [\mathbb{I}_{n+1} - \mathbb{I}_n]\,.
\end{equation}
In this way, we find
\begin{equation}\label{M15}
\mathbb{I}_0 = \frac{\mathbb{E}_0}{(1- \mathbb{E}_0^2)^{3/2}}\,, \quad \mathbb{I}_1 =  \frac{3}{2}\frac{\mathbb{E}_0}{(1- \mathbb{E}_0^2)^{5/2}}\,, \quad \mathbb{I}_2 = \frac{1}{2} \frac{(4 +\mathbb{E}_0^2)\mathbb{E}_0}{(1- \mathbb{E}_0^2)^{7/2}}\,,~~ {\rm etc.}\,
\end{equation}
Substituting these results in Eq.~\eqref{M11}, we find for the rate of pericenter precession 
\begin{equation}\label{M16}
\left \langle \frac{d \mathbb{G}}{dt} \right \rangle  = -\frac{GM}{\lambda_0 a_0} \frac{(1- \mathbb{E}_0^2)^{1/2}}{\mathbb{A}_0\,\Omega}\left[\Delta_2 +\frac{3}{2} \Delta_3 \frac{\mathbb{A}_0}{a_0} + \frac{1}{2} \Delta_4 (4+ \mathbb{E}_0^2)\frac{\mathbb{A}_0^2}{a_0^2} + \cdots\right]\,.
\end{equation}
This general result can also be obtained by studying the average motion of the Runge-Lenz vector, as explained below.

\subsection{Precession of the Runge-Lenz Vector}

For the perturbed Keplerian system~\eqref{M1}, we define the Runge-Lenz vector $\mathbf{Q}$ by
\begin{equation}\label{M17}
\mathbf{Q} := \mathbf{V} \times \mathbf{L} - GM \, \frac{\mathbf{R}}{R}\,,
\end{equation}
where $\mathbf{V} = d\mathbf{R}/dt$ and $\mathbf{L} = \mathbf{R} \times \mathbf{V}$ are the relative velocity and specific orbital angular momentum of the relative motion, respectively. The latter is in the $Z$ direction and has magnitude $L$.   It follows from Eqs.~\eqref{M1} and~\eqref{M17} that
\begin{equation}\label{M18}
\frac{d\mathbf{Q}}{dt} = -\mathbb{F}(R)\,\frac{\mathbf{R}}{R} \times\mathbf{L} = L\,\mathbb{F} \,(-\sin\varphi, \cos\varphi, 0)\,,
\end{equation}
in the background Cartesian system of coordinates. For the unperturbed orbit, $\mathbf{Q}$ is a constant of the motion and is given by 
\begin{equation}\label{M19}
\mathbf{Q}_0 := GM \mathbb{E}_0 \,(\cos\mathbb{G}_0, \sin\mathbb{G}_0, 0)\,.
\end{equation}
The Runge-Lenz vector points in the direction of the pericenter and its magnitude is characterized by the eccentricity of the orbit; indeed, it vanishes for a circular Keplerian orbit. 

For a small perturbing function as in Eq.~\eqref{M1}, the rate of temporal variation of $\mathbf{Q}$ is ``slow"; therefore, its overall precession becomes evident once we average Eq.~\eqref{M18} over the ``fast" Keplerian motion. Employing the same averaging procedure as in the first part of this section, we find 
\begin{equation}\label{M20}
\left \langle \frac{d \mathbf{Q}}{dt} \right \rangle  = \mathbb{Q}\, (- \sin\mathbb{G}_0, \cos\mathbb{G}_0, 0)\,,
\end{equation}
where
\begin{equation}\label{M21}
\mathbb{Q} = - L_0 \sum_{n=0}^{\infty}\mathbb{A}_0^{n} (1-\mathbb{E}_0^2)^{n+3/2}\mathbb{F}_n\,\mathbb{I}_n\,.
\end{equation}
In computing the average in Eq.~\eqref{M20}, we have assumed $\varphi-\mathbb{G}_0 = \eta$ and used the integrals
\begin{equation}\label{M22}
\frac{1}{2\pi} \int_{0}^{2\pi} \frac{\sin \varphi\,d\varphi}{[1+ \mathbb{E}_0 \cos(\varphi-\mathbb{G}_0)]^{n+2}} = - \mathbb{I}_n\,\sin\mathbb{G}_0\,, 
\end{equation}
\begin{equation}\label{M23}
\frac{1}{2\pi} \int_{0}^{2\pi} \frac{\cos \varphi\,d\varphi}{[1+ \mathbb{E}_0 \cos(\varphi-\mathbb{G}_0)]^{n+2}} = - \mathbb{I}_n\,\cos\mathbb{G}_0\,, 
\end{equation} 
where $\mathbb{I}_n$ is given by Eq.~\eqref{M12}. 

Finally, we can write
\begin{equation}\label{M24}
\left \langle \frac{d \mathbf{Q}}{dt} \right \rangle  = \boldsymbol{\omega} \times \mathbf{Q}_0\,,
\end{equation}
where frequency $\boldsymbol{\omega}$ points in the $Z$ direction and its magnitude $\omega$ is the rate of pericenter precession. Therefore,  
\begin{equation}\label{M25}
\omega = \frac{\mathbb{Q}}{GM \mathbb{E}_0} = - \frac{1}{\mathbb{E}_0\Omega} \sum_{n=0}^{\infty}\mathbb{A}_0^{n-1} (1-\mathbb{E}_0^2)^{n+2}\mathbb{F}_n\,\mathbb{I}_n\,,
\end{equation}
in agreement with Eq.~\eqref{M11}.

\section{Improved Lower Bound for $a_0$}

The nonlocal modifications of gravity in the solar system crucially depend on the short-range parameter $a_0$ of the reciprocal kernel. To emphasize this point, let us write the result of the previous section for the perihelion precession frequency of planets beyond the Earth more explicitly as follows:
\begin{align}\label{N1}
\nonumber \omega|_{q = q_1} = {}& -\frac{1}{2} \Omega  \frac{\mathbb{A}_0^2}{\lambda_0 a_0}\,(1- \mathbb{E}_0^2)^{1/2} \Big[1+\zeta_0 - (1+\zeta_0 +\zeta_0^2)\frac{\mathbb{A}_0}{a_0} \\
{}& + (1+\zeta_0 +\tfrac{1}{2}\zeta_0^2 + \tfrac{1}{2}\zeta_0^3) (1+\tfrac{1}{4} \mathbb{E}_0^2)\frac{\mathbb{A}_0^2}{a_0^2} - \cdots \Big]\,,
\end{align}
\begin{align}\label{N2}
\nonumber \omega|_{q = q_2} = {}& -\frac{1}{2} \Omega \frac{\mathbb{A}_0^2}{\lambda_0 a_0}\,(1- \mathbb{E}_0^2)^{1/2} \Big[(1+\zeta_0)\frac{\mathbb{A}_0}{a_0} \\ 
{}& -2 (1+\zeta_0 +\tfrac{1}{2}\zeta_0^2) (1+\tfrac{1}{4} \mathbb{E}_0^2)\frac{\mathbb{A}_0^2}{a_0^2} + \cdots \Big]\,,
\end{align}
where $\Omega$ is the Keplerian frequency of the planetary orbit and $\zeta_0 = a_0\mu_0$. In these expressions, the terms in the square brackets up to and including those proportional to $\mathbb{A}_0/a_0$ agree with the results of previous   work~\cite{BMB, Chicone:2015coa}; however, terms of order $(\mathbb{A}_0/a_0)^2$ and higher are new. 

In principle, solar system observations can be used to limit the magnitude of the short-distance parameter $a_0$~\cite{Iorio:2014roa}. In 2015, observational data regarding the rate of perihelion precession of Saturn was employed to set an important lower bound on $a_0$ of about $10^{15}$ cm; in fact, $a_0 \gtrsim 2 \times 10^{15}$ cm if $q_1$ was used for the reciprocal kernel of NLG in the Newtonian regime, whereas $a_0 \gtrsim \tfrac{1}{2} \times 10^{15}$ cm if  $q_2$ was used~\cite{Chicone:2015coa}.  Specifically, to avoid conflict with high-precision ephemerides, it was assumed that nonlocal gravity should not produce an extra shift in the perihelion precession of Saturn that in absolute magnitude would be greater that $2 \times 10^{-3}$ seconds of arc per century.   To get a significant lower bound on $a_0$ in this way, the planetary orbit should be rather far from the Sun~\cite{Chicone:2015coa}. Indeed, Saturn's orbit has a semimajor axis $\mathbb{A}_0 \approx 9.58$ AU, eccentricity $\mathbb{E}_0\approx 0.0565$ and an orbital period  $T \approx 29.46$ yr.   In the meantime, improved data for the rate of perihelion precession of Saturn has become available~\cite{Iorio:2018adf}. Based on the new data, we can now assume that the absolute magnitude of the extra contribution of nonlocal gravity to the rate of perihelion precession of Saturn must not exceed $0.67 \times 10^{-3}$ seconds of arc per century. This conservative estimate increases by a factor of $3$ the lower limit on $a_0$. That is, if the reciprocal kernel of nonlocal gravity is $q_1$, we find $a_0 \gtrsim 400$ AU. However, if the reciprocal kernel is $q_2$, we find $a_0 \gtrsim 100$ AU. 

\section{Discussion}

This paper goes beyond the preliminary calculations contained in previous work~\cite{BMB, Chicone:2015coa} and presents a general treatment of the NLG deviations from the inverse square law of gravity in the solar system. To obtain explicit expressions for the deviations due to NLG, we have assumed that the Sun has uniform density and is spherically symmetric. These simplifying assumptions in the specific case of classical nonrelativistic nonlocal modification of Newtonian gravity are  adequate at present; however, better approximations may be necessary when detailed observational data become available in the future. In any case, with these assumptions the force of gravity on a point mass $m$ at a distance $R$ from the Sun is radial toward the Sun in NLG and is given by
\begin{equation}\label{D1}
 \frac{GmM_\odot}{R^2}\,\left[1+ \frac{1}{2} (1+\zeta_0)\frac{R^2}{\lambda_0a_0} -\frac{1}{3} (1+\zeta_0 + \zeta_0^2)\frac{R^3}{\lambda_0a_0^2}+\frac{1}{4} (1+\zeta_0 +\tfrac{1}{2} \zeta_0^2 +\tfrac{1}{2} \zeta_0^3)\frac{R^4}{\lambda_0a_0^3}+ \cdots \right]\,
\end{equation}
or
\begin{equation}\label{D2}
 \frac{GmM_\odot}{R^2}\, \left[1+ \frac{1}{3} (1+\zeta_0)\frac{R^3}{\lambda_0a_0^2} - \frac{1}{2} (1+\zeta_0 +\tfrac{1}{2} \zeta_0^2)\frac{R^4}{\lambda_0a_0^3}+ \cdots \right]\,,
\end{equation}
depending upon whether the reciprocal kernel is chosen to be $q_1$ or $q_2$ as given in Eq.~\eqref{I5}. Terms proportional to $R^4$ inside the square brackets are new and go beyond previous results~\cite{BMB, Chicone:2015coa}; moreover, methods developed in this paper make it possible to calculate all of the higher-order terms in these equations as well as their influence on planetary orbits in the solar system. 
Here, $\zeta_0 = \mu_0 a_0$, $\lambda_0 \approx 3\, {\rm kpc}$ and $\mu_0^{-1} \approx 17\, {\rm kpc}$. Furthermore, employing new observational data regarding the perihelion precession rate of Saturn,  we can provide a new lower limit for the short-range parameter $a_0$ of the reciprocal kernel that is an improvement by a  factor of 3 over the old one~\cite{Chicone:2015coa}; that is, 
$a_0 \gtrsim 400$ AU if the reciprocal kernel $q$  is  $q_1$ and $a_0 \gtrsim 100$ AU if it is $q_2$. Let us recall here the requirement of NLG that the reciprocal kernel be integrable as well as square integrable; indeed,  the presence of $a_0$ is necessary to avoid a singularity at $r = 0$ in the reciprocal kernel and thereby satisfy the reasonable mathematical requirements of NLG.

In the Newtonian regime of NLG, the nonlocal aspect of the universal gravitational interaction appears in effect as dark matter. In modern astronomy, dark matter is needed to explain dynamics of galaxies, clusters of galaxies, and structure formation in cosmology.  In the current standard model of cosmology, the energy content of the universe consists of  about 70\% dark energy, about 25\% dark matter and about 5\% visible matter. Most of the matter in the universe is currently thought to be in the form of certain elusive particles of cold dark matter that, despite much effort, have not been directly detected. 
The existence and properties of this cold dark matter have thus far been deduced only through its gravity. On the other hand, it is possible that there is no dark matter at all and the theory needs to be modified on the scales of galaxies and beyond in order to take due account of what appears as dark matter in astronomy and cosmology. A suitably extended theory of gravitation could then account for observational data without any need for dark matter. In this way, modified gravity theories such as NLG have been constructed. Indeed, other approaches to nonlocal gravitation exist in connection with dark matter or dark energy that draw their inspiration from developments in quantum field theory. In one such class of nonlocal models concerned with cosmic acceleration, there is effectively no deviation from general relativity within the solar system~\cite{Park:2012cp, Woodard:2014iga}.   

The results of the present investigation are expected to be of interest in connection with future experiments that will look for the signature of dark matter in the solar system~\cite{Buscaino:2015fya, Feldman:2016pws, Berge:2019zjj, Zwick:2022hbe}; moreover, such experiments may indeed determine the short-range parameter $a_0$ of the reciprocal kernel $q$.

\section*{Acknowledgments}
 
B. M.  is grateful to Lorenzo Iorio for helpful discussions.  The work of M. R. has been supported by the Ferdowsi University of Mashhad.

\appendix

\section{$\rho_D(r)$ for a Spherical Constant Density Object}

Imagine a spherical system with \emph{constant} density of matter up to radius $r_0$. That is, $\rho(r) = \rho_0$ for $r \in [0, r_0]$ and otherwise $\rho(r) = 0$. We use a Cartesian coordinate system with its origin at the center of the sphere. We want to compute $\rho_D$ throughout  space using the convolution formula given in Eq.~\eqref{I3} with  $q = q_1$ or $q=q_2$. Let us note that $q$ and $\rho$ are spherically symmetric; therefore, $\rho_D(\mathbf{x})$ is spherically symmetric as well and is thus only a function of $r = |\mathbf{x}|$. For the sake of simplicity, we use below
\begin{equation}\label{A1} 
\hat{r}  :=  \mu_0 \,r\,, \qquad \hat{r}_0 := \mu_0\,r_0\,. 
\end{equation} 

\subsection{$q=q_1$}

Let us first choose $q = q_1$, which can now be written as
\begin{equation}\label{A2} 
q_1 = \frac{\alpha_0 \mu_0^3}{8\pi}\, \frac{1 + \zeta_{0} +\hat{U}}{\hat{U}(\zeta_{0} + \hat{U})}\,e^{-\hat{U}}\,,  
\end{equation} 
where $\alpha_0 := 2/(\lambda_0\,\mu_0)$, $\zeta_{0} := \mu_0 a_0$ and 
\begin{equation}\label{A3} 
\hat{U} := \mu_0\, |\mathbf{x}-\mathbf{y}| = \mu_0\,(r^2 + y^2 -2ry \cos\theta)^{1/2}\,, \qquad y = |\mathbf{y}|\,.
\end{equation} 
With no loss in generality, we have assumed here that $\mathbf{x}$ is in the $z$ direction and $\theta$ is the polar angle of $\mathbf{y}$. Therefore, 
\begin{equation}\label{A4} 
^{(1)}\rho_D(r)= \frac{\alpha_0 \rho_0}{4}\,\int \frac{1 + \zeta_{0} +\hat{U}}{\hat{U}(\zeta_{0} + \hat{U})}\,e^{-\hat{U}} \,\hat{y}^2\,d\hat{y} \sin\theta\,d\theta\,,
\end{equation} 
where $\hat{y} := \mu_0 \,y$. From  the integration over $\theta$, we find
\begin{equation}\label{A5} 
 \int \left(1+\frac{1}{\zeta_{0} + \hat{U}}\right)\,e^{-\hat{U}} d\hat{U} = -e^{\zeta_{0}}\left[e^{-\zeta_{0} - \hat{U}} + E_1 (\zeta_{0} + \hat{U})\right]\,.
\end{equation} 
In this way, we find our main result, namely, 
\begin{equation}\label{A6} 
^{(1)}\rho_D(r)= \frac{\alpha_0 \rho_0\,e^{\zeta_0}}{4\,\hat{r}}\int_0^{\hat{r}_0}\hat{y}\,d\hat{y} \left[ e^{-|\hat{r}-\hat{y}|-\zeta_{0}} - e^{-\hat{r}-\hat{y}-\zeta_{0}} + E_1(|\hat{r}-\hat{y}|+\zeta_{0}) - E_1(\hat{r}+\hat{y}+\zeta_{0})\right]\,,
\end{equation} 
where $\hat{U}(\theta = 0) = |\hat{r}-\hat{y}|$ and $\hat{U}(\theta = \pi) = \hat{r}+\hat{y}$. 

To do the integral, the following results are useful:
\begin{equation}\label{A7} 
\int x \sinh x\,dx = x \cosh x - \sinh x\,,  \qquad \int x e^{-x} \,dx = -(1+x) e^{-x}\,.
\end{equation} 
Furthermore, consider
\begin{equation}\label{A8} 
I(\hat{r}_0, A) := \int_0^{\hat{r}_0} \hat{y}\,d\hat{y} E_1(\hat{y}+A)\,, \qquad  I( -\hat{r}_0, A) = \int_0^{\hat{r}_0} \hat{y}\,d\hat{y} E_1(-\hat{y}+A)\,,
\end{equation} 
where $A$ is a constant. Using integration by parts, we find
\begin{equation}\label{A9} 
I(\hat{r}_0, A) = \frac{1}{2} \hat{r}_0^2 E_1(\hat{r}_0 + A) + \frac{1}{2}\int_0^{\hat{r}_0} \frac{\hat{y}^2}{\hat{y}+A} e^{-\hat{y} - A}\,d\hat{y} \,.
\end{equation} 
The end result is
\begin{equation}\label{A10} 
I(\hat{r}_0, A) = \frac{1}{2}(1-A)e^{-A} + \frac{1}{2}A^2 E_1(A) - \frac{1}{2} (1+\hat{r}_0 -A) e^{-\hat{r}_0 - A} + \frac{1}{2} (\hat{r}_0^2-A^2) E_1(\hat{r}_0 + A)\,.
\end{equation} 

Let us first calculate the outside density, namely, $r > r_0$. We have from Eqs.~\eqref{A6}--\eqref{A10},
\begin{equation}\label{A11} 
^{(1)}\rho_D(r>r_0)= \frac{\alpha_0 \rho_0\,e^{\zeta_0}}{4\,\hat{r}}\left[2e^{-\hat{r}-\zeta_{0}}(\hat{r}_0\cosh \hat{r}_0 -\sinh \hat{r}_0) + I(-\hat{r}_0, \hat{r}+\zeta_{0})-I(\hat{r}_0, \hat{r}+\zeta_{0})\right]\,,
\end{equation} 
which works out to be
\begin{align}\label{A12} 
\nonumber  ^{(1)}\rho_D(r>r_0)= {}&\frac{\alpha_0 \rho_0\,e^{\zeta_0}}{8\,\hat{r}}\big\{2e^{-\hat{r}-\zeta_{0}}[3\hat{r}_0\cosh \hat{r}_0 +(\hat{r}+\zeta_{0} -3)\sinh \hat{r}_0]  \\
{}&+ [(\hat{r}+\zeta_{0})^2 - \hat{r}_0^2][E_1(\hat{r}+\zeta_{0} +\hat{r}_0) - E_1(\hat{r}+\zeta_{0}-\hat{r}_0)]\big\}\,.
\end{align} 

Next, we calculate the effective dark matter density in the interior of the compact sphere, i.e. for $r<r_0$. In employing Eq.~\eqref{A6}, we need to split the integration of $y\in (0, r_0)$ into two parts: integration over $y\in (0, r)$, where $r-y>0$, and $y\in (r, r_0)$, where $r-y<0$. Hence, 
\begin{align}\label{A13} 
\nonumber  {}&^{(1)}\rho_D(r<r_0)= \frac{\alpha_0 \rho_0\,e^{\zeta_0}}{4\,\hat{r}}\int_0^{\hat{r}}\hat{y}\,d\hat{y} \left[ e^{-\hat{r}+\hat{y}-\zeta_{0}} - e^{-\hat{r}-\hat{y}-\zeta_{0}} + E_1(\hat{r}-\hat{y}+\zeta_{0}) - E_1(\hat{r}+\hat{y}+\zeta_{0})\right] \\
{}& +\frac{\alpha_0 \rho_0\,e^{\zeta_0}}{4\,\hat{r}}\int_{\hat{r}}^{\hat{r}_0}\hat{y}\,d\hat{y} \left[e^{\hat{r}-\hat{y}-\zeta_{0}} - e^{-\hat{r}-\hat{y}-\zeta_{0}} + E_1(-\hat{r}+\hat{y}+\zeta_{0}) - E_1(\hat{r}+\hat{y}+\zeta_{0})\right]\,.
\end{align} 
We can write this expression as
\begin{align}\label{A14} 
\nonumber ^{(1)}\rho_D(r< r_0)= {}&\frac{\alpha_0 \rho_0}{2\,\hat{r}}[\hat{r}-(1+\hat{r}_0)\,e^{-\hat{r}_0}\,\sinh \hat{r}] + \frac{\alpha_0 \rho_0\,e^{\zeta_0}}{4\,\hat{r}}\,[I(-\hat{r}, \hat{r}+\zeta_{0})\\
{}&  - I(\hat{r}_0, \hat{r}+\zeta_{0}) - I(\hat{r}, -\hat{r}+\zeta_{0}) + I(\hat{r}_0, -\hat{r}+\zeta_{0})]\,.
\end{align} 
Using Eq.~\eqref{A10}, we find after some algebra
\begin{equation}\label{A15} 
^{(1)}\rho_D(r<r_0)= \mathcal{A}_1 + \mathcal{B}_1 + \mathcal{C}_1\,,
\end{equation}
\begin{equation}\label{A16} 
 \mathcal{A}_1  = \alpha_0 \rho_0 [1 - \tfrac{1}{2}\zeta_{0} e^{\zeta_0} E_1(\zeta_{0})]\,,
\end{equation}
\begin{equation}\label{A17} 
 \mathcal{B}_1  = -\frac{\alpha_0 \rho_0}{4\,\hat{r}} e^{-\hat{r}_0}[\hat{r}\cosh \hat{r} + (3+ 3\hat{r}_0 -\zeta_{0}) \sinh \hat{r}]\,,
\end{equation}
\begin{equation}\label{A18} 
 \mathcal{C}_1  = \frac{\alpha_0 \rho_0\,e^{\zeta_0}}{8\,\hat{r}}\{[(\hat{r}+\zeta_{0})^2-\hat{r}_0^2]E_1(\hat{r}_0 +\hat{r} +\zeta_{0}) - [(\hat{r}-\zeta_{0})^2-\hat{r}_0^2]E_1(\hat{r}_0 -\hat{r} +\zeta_{0})\}\,.
\end{equation}

The effective dark matter density in NLG is the convolution of the reciprocal kernel $q$ with the density of matter $\rho$; in the case under consideration here, we have a constant density sphere with radius $r_0$. For reciprocal kernel $q_1$, we have calculated the dark matter density outside $^{(1)}\rho_D(r>r_0)$ and inside $^{(1)}\rho_D(r<r_0)$ the sphere. It is possible to show that these agree on the surface of the sphere $ r = r_0$ and  $^{(1)}\rho_D$ is therefore a continuous function throughout space. 

\subsection{$q=q_2$}

The reciprocal kernel is now $q_2 = [\hat{U}/(\zeta_{0} + \hat{U})] q_1$. Thus, with the same notation and convention as before, we have 
\begin{equation}\label{A19} 
^{(2)}\rho_D(r)= \frac{\alpha_0 \rho_0}{4}\,\int \frac{1 + \zeta_{0} +\hat{U}}{(\zeta_{0} + \hat{U})^2}\,e^{-\hat{U}} \,\hat{y}^2\,d\hat{y} \sin\theta\,d\theta\,.
\end{equation}  
After the $\theta$ integration, namely, 
\begin{equation}\label{A20} 
 \int \frac{(1 + \zeta_{0} +\hat{U}) \hat{U}}{(\zeta_{0} + \hat{U})^2}\,e^{-\hat{U}} d\hat{U} =  - \frac{\hat{U}}{\zeta_{0} + \hat{U}} e^{-\hat{U}} -e^{\zeta_{0}} E_1 (\zeta_{0} + \hat{U})\,,
\end{equation} 
we find
\begin{align}\label{A21} 
\nonumber   ^{(2)}\rho_D(r)= {}&\frac{\alpha_0 \rho_0}{4\,\hat{r}}\int_0^{\hat{r}_0}\hat{y}\,d\hat{y} \left( \frac{|\hat{r}-\hat{y}|}{|\hat{r}-\hat{y}|+\zeta_{0}}\,e^{-|\hat{r}-\hat{y}|} - \frac{\hat{r}+\hat{y}}{\hat{r}+\hat{y}+\zeta_{0}}\,e^{-\hat{r}-\hat{y}}\right)  \\
{}& +\frac{\alpha_0 \rho_0\,e^{\zeta_0}}{4\,\hat{r}}\int_0^{\hat{r}_0}\hat{y}\,d\hat{y} [E_1(|\hat{r}-\hat{y}|+\zeta_{0}) - E_1(\hat{r}+\hat{y}+\zeta_{0})]\,,
\end{align} 
where $\hat{U}(\theta = 0) = |\hat{r}-\hat{y}|$ and $\hat{U}(\theta = \pi) = \hat{r}+\hat{y}$. 

For the outside region ($r > r_0$), we use the methods described above to find
\begin{align}\label{A22} 
\nonumber  ^{(2)}\rho_D(r>r_0)= {}&\frac{\alpha_0 \rho_0\,e^{\zeta_0}}{8\,\hat{r}}\big\{2e^{-\hat{r}-\zeta_{0}}[3\hat{r}_0\cosh \hat{r}_0 +(\hat{r}+ 3\zeta_{0} -3)\sinh \hat{r}_0]  \\
{}&+ [(\hat{r}+\zeta_{0})(\hat{r} + 3 \zeta_{0}) - \hat{r}_0^2][E_1(\hat{r}+\zeta_{0} +\hat{r}_0) - E_1(\hat{r}+\zeta_{0}-\hat{r}_0)]\big\}\,.
\end{align} 
In a similar way as before, we find the effective density of dark matter inside the spherical object of constant density $\rho_0$, namely, 
\begin{equation}\label{A23} 
^{(2)}\rho_D(r<r_0)= \mathcal{A}_2 + \mathcal{B}_2 + \mathcal{C}_2\,,
\end{equation}
\begin{equation}\label{A24} 
 \mathcal{A}_2  = \alpha_0 \rho_0 [1 - \zeta_{0} e^{\zeta_0} E_1(\zeta_{0})]\,,
\end{equation}
\begin{equation}\label{A25} 
 \mathcal{B}_2  = -\frac{\alpha_0 \rho_0}{4\,\hat{r}} e^{-\hat{r}_0}[\hat{r}\cosh \hat{r} + 3(1+ \hat{r}_0 -\zeta_{0}) \sinh \hat{r}]\,,
\end{equation}
\begin{equation}\label{A26} 
 \mathcal{C}_2  = \frac{\alpha_0 \rho_0\,e^{\zeta_0}}{8\,\hat{r}}\{[(\hat{r}+\zeta_{0})(\hat{r}+3\zeta_{0})-\hat{r}_0^2]E_1(\hat{r}_0 +\hat{r} +\zeta_{0}) - [(\hat{r}-\zeta_{0})(\hat{r}-3\zeta_{0})-\hat{r}_0^2]E_1(\hat{r}_0 -\hat{r} +\zeta_{0})\}\,.
\end{equation}

\begin{figure}
\centering
\includegraphics[width=8.0cm]{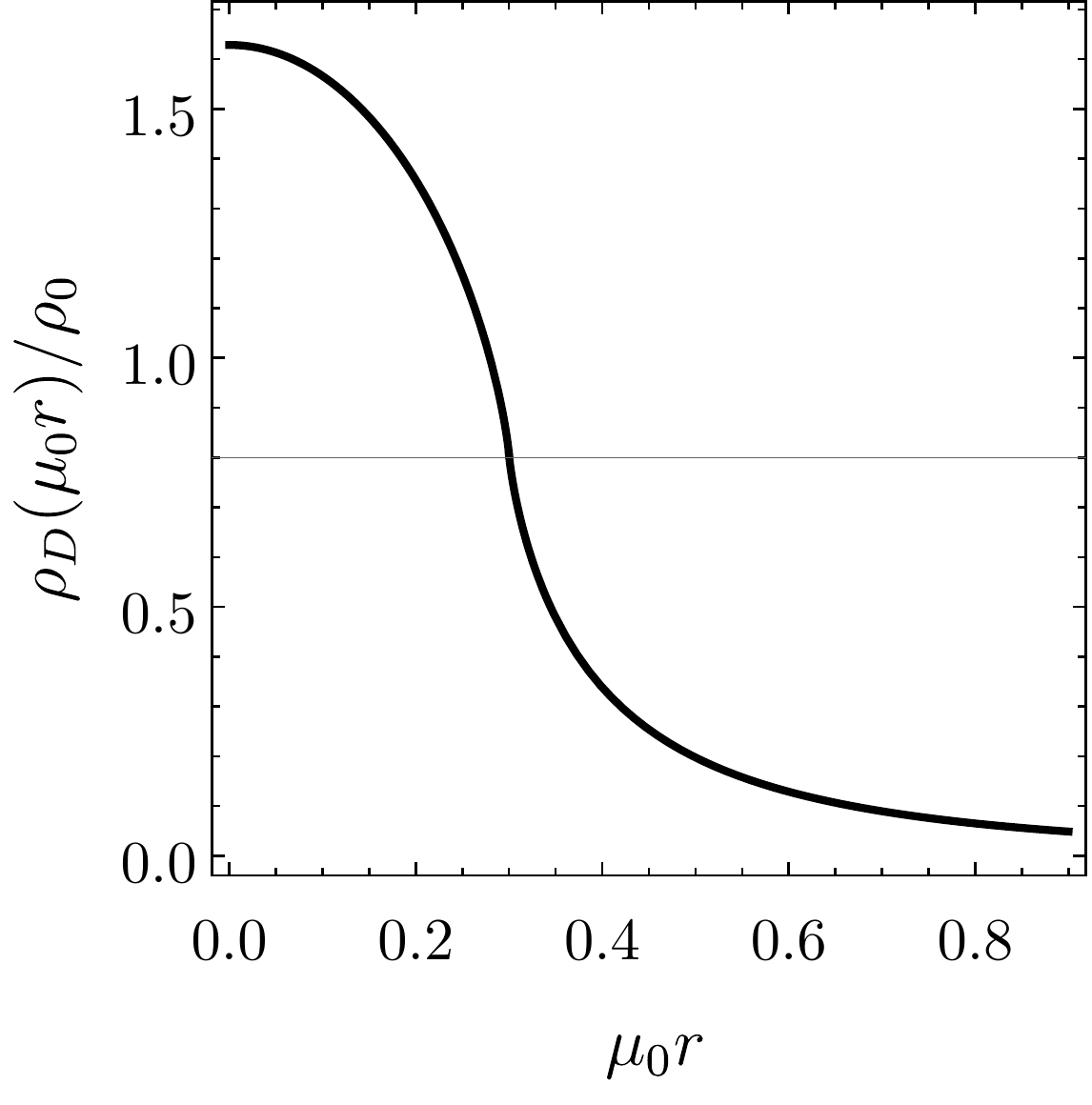}
\caption{The effective dark matter for a constant density sphere given by Eqs.~\eqref{A22} and~\eqref{A23}. Here, $\rho_D(\hat{r})/\rho_0$ is plotted versus $\hat{r} = \mu_0 r$ and we have assumed that $\zeta_{0}=0$ and $\hat{r}_0 = \mu_0r_0=0.3$. }\label{con_den}
\end{figure}

On physical grounds, we expect that $\zeta_{0} < 1$; for instance, a reasonable value for the short distance parameter $a_0$ is $10^{18}$\,cm $\approx 1$ light-year, in which case $\zeta_{0} \approx 2 \times 10^{-5}$ is very small compared to unity and one cannot observationally distinguish the two cases involving $q_1$ and $q_2$. For $\zeta_{0} < 1$, $\rho_D(r)$ has a maximum with zero slope at the center of the object ($r = 0$). For $\zeta_{0} = 0$, the two cases agree, since $q_1=q_2 = q_0$; in this case, Figure~\ref{con_den} illustrates the behavior of $\rho_D(r)$ for $\hat{r}_0 = \mu_0 r_0 = 0.3$.

\end{document}